# Spin-dependent transport of electrons in the presence of smooth lateral barrier and spin-orbit interaction

Alexander O. Govorov[1], Alexander V. Kalameitsev[2], and John P. Dulka[1]

[1]Department of Physics and Astronomy, Ohio University, Athens OH 45701

[2]Institute of Semiconductor Physics, Novosibirsk, 63090, Russia

**Abstract**

We describe theoretically the process of multi-beam reflection in a two-dimensional electron system with a lateral potential barrier. Due to spin-orbital interaction, the reflection process leads to the formation of three beams with different spin polarizations. The efficiency of spin conversion can become small for smooth lateral barriers. Nevertheless, we demonstrate that the spin-conversion effect remains strong for realistic lithographical potentials and spin-orbit interactions in etched lateral nano-structures. The system with a lateral barrier suggests useful applications as a spin-filtering device. The expected quasi-classical adiabatic behavior without spin conversion is found in the system with a very strong spin-orbit interaction. We also consider the quasi-classical motion of electrons in a system with boundaries in a magnetic field and two magnetic focusing geometries.

PACS: 73.23.Ad, 72.25.Dc

## Introduction

Mobile electrons in meso-scopic and nano-scopic structures experience the spin-orbit interaction (SOI) and therefore the translational and spin motions of an electron become coupled. The most common type of measurements in semiconductor nanostructures concerns electric currents. However, since the SOI in the most common semiconductors is weak, electric-current measurements can be relatively insensitive to spins. There are several methods to enhance the spin-effects in the electric-current measurements, such as the use of ferromagnetic leads to the semiconductor quantum well [1], driving currents through multi-barrier structures, waveguides, channels, or quantum dots [2], etc.

In the absence of electric and magnetic fields, electrons with opposite spins move along a straight line. However, when electric and magnetic fields are present, electron trajectories depend on a spin state and the SOI-effects in the electric conductance become enhanced. Recently, several mechanisms of spatial separation of electron beams with different spin orientations in ballistic lateral nano-structures have been proposed. The spin-polarized beams can be obtained by using spin-dependent reflection from a lateral barrier [3], a spatially varied SOI interaction [4], and cyclotron motion [5]. In this paper, we focus on physical properties of spin-dependent reflection in a two-dimensional (2D) electron system with a lateral barrier, following the method proposed in ref. [3].

Here we describe the electron reflection process in a 2D gas with a SOI. This process has a multi-beam character and can be utilized for spin filtering in meso-scopic 2D systems with ballistic electron beams figs. 1 a) and b). In such structures, an electron beam injected from the incoming window becomes reflected from a lateral barrier, and then leaves the system through the outgoing aperture. A weak magnetic field serves as a tool to focus the electron beam to the outgoing aperture. Due to the SOI, the incoming beam becomes split into three beams at the lateral barrier. This effect comes from the simple kinematics and will be explained below. In the paper, we show that the effect of multi-beam reflection is strong if the lateral barrier is sharp enough. In the case of a very smooth potential, the spin-dependent reflection vanishes. Here we calculate the reflection coefficient for etched semiconductor nano-structures and show that the multi-beam reflection effect is strong for realistic lateral barriers with relatively smooth potential profiles.

## 1. Model

The motion of single electrons in a 2D system in the presence of electric and magnetic fields is described by the Hamiltonian:

$$\hat{H} = \frac{\hat{P}_x^2 + \hat{P}_y^2}{2m_e} + U(\vec{r}) + \hat{V}_{SO}, \qquad (1)$$

where $\vec{r} = (x, y)$ is the in-plane radius vector, $U(\vec{r}) = e\varphi(\vec{r})$, $\hat{\vec{P}} = \hat{\vec{p}} - \frac{e}{c}\vec{A}$, and $\hat{\vec{p}}$ is the in-plane momentum operator; $\vec{A}$ and $\varphi$ are the vector and scalar potentials, respectively; $\hat{V}_{SO}$ represents the SOI. For the SOI, we assume the Bychkov-Rashba inversion asymmetry mechanism induced by in-plane and perpendicular electric fields [6].

We now consider the reflection process for a one-dimensional barrier $U(x) = e\varphi(x)$ in an asymmetric quantum well. In this case, the SOI-operator takes a form,

$$\hat{V}_{SO}(x) = \frac{\alpha_{SO}}{\hbar}(\hat{\sigma}_x \hat{P}_y - \hat{\sigma}_y \hat{P}_x) + \frac{\gamma}{\hbar}\frac{dU(x)}{dx}\hat{\sigma}_z \hat{P}_y , \qquad (2)$$

where $\hat{\sigma}_i$ are the Pauli matrixes. The first term in eq. 2 originates from the perpendicular electric fields in an asymmetric quantum well, and the second from the in-plane electric field. The material parameter $\gamma$ describes the strength of the SOI. If the SOI arises from the lateral and perpendicular build-in electric fields, the parameter $\alpha_{SO} = -e\gamma F_z$, where $F_z$ is the strength of the perpendicular build-in electric field. The operator (2) is written for the lowest 2D subband after averaging over the $z$-direction.

First we consider single-electron wave function of in the absence of the fields. According to eq. 1, the wave functions and energy of a single electron have a form:

$$\Psi_{\vec{k},\pm} = \frac{1}{\sqrt{2}}\begin{pmatrix} 1 \\ \pm e^{i\varphi(\vec{k})}/i \end{pmatrix} e^{i\vec{k}\vec{r}}, \quad E_{\vec{k},\pm} = \frac{\hbar^2 k^2}{2m_e} \pm \alpha_{so} k , \qquad (3)$$

where $\vec{k} = (k_x, k_y)$ and $\tan\varphi(\vec{k}) = k_y / k_x$. In the above states $\pm$, the spin is perpendicular to the momentum due to the SOI.

## 2. The kinematics of reflection at zero magnetic field

The kinematics of the reflection process is shown in figs. 2 b) and c) [3]. Electrons are injected with the Fermi energy $E_F$. According to the equations (3), electrons with different spin orientations have different momentums (fig. 2 a):

$$k_{F,\pm} = k_F \sqrt{1 + \frac{\delta^2}{16} \mp \frac{\alpha m_e}{\hbar^2}},$$

where $k_F$ is the characteristic Fermi momentum, $E_F = \frac{\hbar^2 k_F^2}{2m_e}$, the SOI dimensionless parameter $\delta = \frac{\Delta_{SO}}{E_F}$; here $\Delta_{SO} = 2\alpha_{SO} k_F$ is the spin splitting at the Fermi energy.

In the geometry of fig. 2b, electrons are injected at the incident angle $\theta$ with respect to the normal. The incident $\Psi_+$ wave turns into two beams with $\pm$ spin configuration (see inserts in the fig. 2c). Similarly, the $\Psi_-$ wave also creates two scattered beams. If the incoming beam is not spin polarized and composed of $\pm$ electrons, the resulting scattered wave is composed of three beams. The central beam is not polarized, whereas the side beams are fully spin polarized. For a given incident angle $\theta$, the components of the momentum for the incoming wave are obvious, $k_{x,\pm} = -k_{F,\pm}\cos(\theta)$ and $k_{y,\pm} = -k_{F,\pm}\sin(\theta)$. In the reflection process, the electron conserves its energy and $k_y$ and, therefore, the wave function can be written as $\Psi = \Phi(x)e^{ik_y y} = \Psi^{in} + \Psi^{out}$. The incoming wave is assumed to be pure, $\Psi^{in} = \Psi_{\vec{k}_\pm,\pm}$, where $\vec{k}_\pm = (k_{x,\pm}, k_{y,\pm})$, $\tan(\theta) = k_{x,\pm}/k_{y,\pm}$, and $E_{\vec{k}_\pm,\pm} = E_F$. Since the spin state is not conserved in the reflection process, the reflected wave has two components propagating at different angles: $\Psi^{out} = A(+)\Psi_{\mathbf{q}_+,+} + A(-)\Psi_{\mathbf{q}_-,-}$, where the momentums in the reflected waves, $\vec{q}_+ = (q_{x+}, k_y)$ and $\vec{q}_- = (q_{x-}, k_y)$, are determined by the kinematics conditions sketched in fig. 2c (for details see ref. [3]). In fig. 2c, we also show the scattering angles $\theta_{+\to-}$ and $\theta_{-\to+}$ for the processes $+\to-$ and $-\to+$. The processes $+\to+$ and $-\to-$ conserves the angle: $\theta_{-\to-} = \theta_{+\to+} = \theta$. Note that the incident wave $\Psi_-^{in}$ has a critical angle $\theta_c$ at which the second scattered beam vanishes; in the limit $\sqrt{\delta} \ll 1$, we obtain $\theta_c \approx \pi/2 - \sqrt{\delta}$. For the angles $\theta > \theta_c$, the electron wave function contains a wave localized nearby the barrier and propagating in the $-y$ direction. Above, we used the typical parameters of InSb quantum wells, $m_e = 0.014\,m_0$, $\alpha_{SO} = 10^{-6}\,meV\,cm$, and $\gamma = 1\cdot 10^{-14}\,cm^2$ [7].

### 3. Quasi-classical motion

In the presence of weak and smooth fields, the motion of an electron is quasi-classical and is given by the usual equations:

$$\hbar\dot{\vec{k}}_\alpha = \vec{F}, \qquad \vec{v}_\alpha = \frac{1}{\hbar}\frac{\partial E_\alpha}{\partial \vec{k}_\alpha}, \qquad (4)$$

where $\vec{F}$ is the classical force and $\alpha = \pm$ is the spin-state index. This approximation implies that the electron does not make transitions between the spin states $\alpha = \pm$.

It is easy to solve the above quasi-classical equations in the uniform magnetic field $B$. This solution will allow us to analyze the electron motion outside the barrier. In the presence of SOI, the cyclotron radii for the states $\pm$ at the Fermi level become slightly different, $R_{c,\pm} = \hbar k_{F,\pm} / m_e \omega_c$, where $k_{F,\pm}$ are the Fermi wave vectors and $\omega_c = |e|B / m_e c$. If the magnetic field is weak enough, it does not affect the reflection process. In this case, the scattering angles $\theta_{\alpha \to \alpha'}$ are given by the above kinematics equations while the motion outside the barrier region is described by eqs. 4. Figure 1 shows the calculated trajectories for the two geometries. In the first case, the system contains two windows, incoming and outgoing. The lateral dimension of this structure is relatively small. The weak magnetic field is used only for focusing. Spatial separation of the ± beams occurs mostly due to reflection because the magnetic field needed for the focusing in this structure is relatively weak. Experimentally, this system should be made as open as possible to avoid additional geometrical resonances [8]. As the magnetic field increases, a different number of reflected beams can pass through the outgoing window and therefore the magnetoresistance of this structure can show the presence of three beams [3]. The second case in fig. 1 is the well-known magnetic focusing geometry explored in the past [9]. In this case, different spin states/beams can be spatially separated due to both the cyclotron motion [5] and spin-dependent reflection.

**4. Reflection from smooth and sharp barriers in weak magnetic fields**

The quasi-classical equations (4) become invalid in the vicinity of classical turning points where the quantum description is necessary. If the electric field of the barrier is strong enough, an electron makes transitions between different spin states. To illustrate the reflection process, we now consider a barrier with a nonzero lateral electric field $F_x(x)$ in the region $-a/2 < x < a/2$. For the potential, we choose $U(x) = U_0(\frac{1}{2} - \frac{3x}{2a} + \frac{2x^3}{a^3})$ if $-a/2 < x < a/2$, $U(x) = 0$ if $x > a/2$, and $U(x) = U_0$ if $x < -a/2$ (fig. 3d). For the above potential, the lateral electric field $F_x$ is about $U_0/|e|a$. The magnetic field is assumed to be weak, so that it does not influence the electron motion in the vicinity of the barrier ($a << R_c$).

If a reflecting barrier is smooth, we can analyze the electron motion in the spirit of the physics of metals [10]. In other words, we can consider the two spin-split sub-bands in a 2D system as two bands in a crystal. In fig. 3, we show qualitatively electron trajectories for different incident angles in the case of a smooth, adiabatic potential, $eF_x\lambda_F << \Delta_{SO}$, where $\Delta_{SO}$ is the spin splitting at the Fermi energy and $\lambda_F = 2\pi/k_F$. The energy $eF_x\lambda_F$ is a characteristic energy of mixing between two spin-split sub-bands. For the incident angle $\theta = 0$, the motion cannot be described by the quasi-classical equations in the vicinity of the classical turning point. In this case, the incident wave $\Psi_+^{in}$ is always converted into the state $\Psi_-^{out}$. This follows from the fact that the spin and translational motions in the Hamiltonian (1) are separated for $\theta = 0$. The quasi-classical approach is applicable to a given point of trajectory if the SOI-splitting at this point $\Delta_{SO} = 2\alpha_{SO}k$ is larger than $eF_x\lambda = eF_x 2\pi/k$. The minimum spin-splitting along a trajectory corresponds to the classical turning point ($k_x = 0$) and is given by $\Delta_{SO,\min} = 2\alpha_{SO}k_y$. An entire trajectory can be described with the quasi-classical approach for sufficiently large incident angles when $\Delta_{SO,\min} = 2\alpha_{SO}k_y > eF_x 2\pi/k_y$. If the SOI-splitting $\Delta_{SO}$ at some point of trajectory is comparable with the electric field energy $eF_x\lambda = eF_x 2\pi/k$, the incident beam $\Psi_+^{in}$ is partially converted into $\Psi_-^{out}$ and vise versa. In the regions with a strong SOI splitting $(\Delta_{SO} = 2\alpha_{SO}k > eF_x\lambda)$, the wave function can be written in a quasi-classical fashion. To write down analytical equations for the wave functions we now neglect the last term in the spin operator (2), assuming a smooth lateral potential. For the incoming + wave, we obtain:

$$\Psi = \Psi_+^{in} + \Psi^{out}, \qquad (5)$$

$$\Psi_+^{in} = \frac{A_{in}}{\sqrt{2v_{x,+}(x)}} \begin{pmatrix} 1 \\ e^{i\varphi_+^{in}(x)}/i \end{pmatrix} e^{ik_y y - \int^x k_{x,+}(x)dx},$$

$$\Psi^{out} = \frac{A_+(+)}{\sqrt{2v_{x,+}(x)}} \begin{pmatrix} 1 \\ e^{i\varphi_+^{out}(x)}/i \end{pmatrix} e^{ik_y y + \int^x k_{x,+}(x)dx} + \frac{A_+(-)}{\sqrt{2v_{x,-}(x)}} \begin{pmatrix} 1 \\ -e^{i\varphi_-^{out}(x)}/i \end{pmatrix} e^{ik_y y + \int^x k_{x,-}(x)dx},$$

where the wave vectors $k_{x,\pm}(x) > 0$ are given by the equation $E_{k_x(x),k_y,\pm} = E_F$. The y-component of the momentum is conserved and $k_y < 1$; $v_{x,\pm}(x) = (1/\hbar)\partial E_\pm/\partial k_x |_{k_x=k_{x,\pm}(x)}$, $|A_{in}| = 1$, $\tan\varphi_+^{in}(x) = |k_y|/k_{x,+}(x)$ and $\tan\varphi_\pm^{out}(x) = -|k_y|/k_{x,\pm}(x)$. The amplitudes in eq.5 obey the conservation law: $\cos(\theta)|A_{in}|^2 = \cos(\theta)|A_+(+)|^2 + \cos(\theta_{+\to-})|A_+(-)|^2$; the lower index in $A_+(\pm)$ denotes a type of incoming wave. For the x-component of the momentum, we have

$$k_{x,\pm}(x) = \sqrt{m^2\left[\sqrt{\alpha^2 + 2/m_e(E_F - U(x))} \mp \alpha\right]^2 - k_y^2}.$$

We now consider the case of entirely quasi-classical trajectory in a sense of the inequality $\Delta_{SO,\min} = 2\alpha_{SO}k_y > eF_x 2\pi/k_y$; this inequality implies $k_y \neq 0$. Then, $A_+(-) \approx 0$ in eq. 5 and the outgoing wave remains in the state + (fig. 3c). The equations (5) diverge at the classical turning point where $k_{x,+}(x) = 0$ and $v_{x,+}(x) = 0$. Near the classical turning point for the + wave, $x_{0+}$, the Hamiltonian can be approximated as

$$\hat{H} \approx -\frac{\hbar^2}{2m_e}\frac{\partial^2}{\partial x^2} + \frac{\hbar^2 k_y^2}{2m_e} + U(x_{0+}) - eF_0(x - x_{0+}) + \alpha_{SO}\hat{\sigma}_x k_y, \qquad (6)$$

where $F_0 = F_x(x_{0+}) = -e^{-1}\partial U/\partial x$. The spin and translational variables in eq. 6 can be separated. The solution has the well know form: $\Psi = const \cdot Ai(\frac{x - x_{0+}}{l_0})$, where $Ai$ is the Airy function and $l_0 = (\hbar^2/2m_e|F_0|)^{1/3}$. Then, by using the asymptotic behavior of the Airy function, we obtain

$$\Psi_+^{in} = \frac{1}{\sqrt{2v_{x,+}(x)}}\begin{pmatrix} 1 \\ e^{i\varphi_+^{in}(x)}/i \end{pmatrix} e^{ik_y y - \int_{x_{0+}}^{x} k_{x,+}(x)dx + \frac{\pi}{4}}, \qquad (7)$$

$$\Psi_+^{out} = \frac{1}{\sqrt{2v_{x,+}(x)}}\begin{pmatrix} 1 \\ e^{i\varphi_+^{out}(x)}/i \end{pmatrix} e^{ik_y y + \int_{x_{0+}}^{x} k_{x,+}(x)dx + \frac{\pi}{4}}.$$

On the trajectory described by eqs. (7), the electron spin follows the orbital motion and is always perpendicular to the momentum $\vec{k} = (k_x(x), k_y)$, as shown in fig. 3 c.

To conclude this section, we note that the trajectory (a) in fig. 3 can be easily described within the quasi-classical approach as the spin and the coordinate $x$ can be separated. In the case of the trajectory (b), the incoming + wave creates two outgoing waves. This case should be treated separately introducing slow conversion of waves in the barrier region [11]. Another solvable limit is $a \to 0$. In this case, the barrier becomes rectangular and we can analytically solve the problem of reflection using the plane waves and exponential functions [3].

## 5. Numerical results

Now we present numerical results that support the above quasi-classical consideration. The barrier potential $U(x)$ has been specified above. Again we use the typical parameters of InSb quantum wells [8]. Experimentally, the conventional methods to fabricate lateral barriers are etching of surface or deposition of a metallic gate; with the above methods, the typical lateral dimensions for the barrier potential cannot be made too short. Typically, they are in the sub-μm range. Figure 4 shows calculated reflection coefficients for the barriers with different widths $a$. The reflection coefficients are defined in the following way:

$$R_{+\to+} = \left|\frac{A_+(+)}{A_{in}}\right|^2, \quad R_{+\to-} = \frac{\cos\theta_{+\to-}}{\cos\theta}\left|\frac{A_+(-)}{A_{in}}\right|^2,$$

$$R_{-\to-} = \left|\frac{A_-(-)}{A_{in}}\right|^2, \quad R_{-\to+} = \frac{\cos\theta_{-\to+}}{\cos\theta}\left|\frac{A_-(+)}{A_{in}}\right|^2.$$

where $A_\pm(\pm)$ are the coefficients of the wave function outside of the barrier, in the right hand side of the system. The above coefficients satisfy the conservation-of-charge law: $R_{+\to+} + R_{+\to-} = 1$ and $R_{-\to-} + R_{-\to+} = 1$. We obtained these equations considering the *x*-component of the current operator. It is seen from fig. 4a that, for a moderate SOI in the InSb system, the reflection coefficients weakly depend on the barrier width and the spin conversion remains very strong even for very smooth lateral barriers. For the barrier $a = 100\,\text{Å}$, the reflection coefficients are very close to those of the hard reflecting wall. For the hard wall barrier, we can use a simple geometrical consideration for the spin conversion at the barrier and obtain: $R_{--} = R_{++} = [1 - \cos(2\theta)]/2$ and $R_{-+} = R_{+-} = [1 + \cos(2\theta)]/2$ [3]. These simple equations are valid if $\sqrt{\delta} \ll 1$ and $\pi/2 - \theta \gg \sqrt{\delta}$. It is also seen from fig. 4a that the off-diagonal coefficients and the spin-conversion effect decrease with increasing the barrier width *a*. It is expected from the quasi-classical theory. The off-diagonal coefficients ($R_{+\to-}$ and $R_{-\to+}$) do not decrease much with increasing *a* because the parameter of the quasi-classical theory ($\eta = \dfrac{eF_0 \lambda_F}{\Delta_{SO}} \approx \dfrac{U_0 \lambda_F / a}{\Delta_{SO}}$) remains large even for the longest barrier. For the longest barrier width of $10000\,\text{Å}$ and $U_0 = 150\,meV$, $\eta \approx 3.6$. This reflects also a relatively small strength of SOI in the conduction band of InSb. Suppression of spin-flip processes for reflection from the smoother potential can be seen in fig. 4b, which shows the reflection coefficients for the barrier of $U_0 = 70\,meV$ ($\eta \approx 1.8$). The figure 5 demonstrates $R_{\alpha,\alpha'}$ for different parameters $\alpha_{SO}$ and for a sharp barrier with $a = 100\,\text{Å}$. The very strong effect coming from $\alpha_{SO}$ is seen in reflection of the wave $\Psi_-^{in}$; this effect originates from the kinematics of scattering. As for the reflection coefficients $R_{++}$ and $R_{+-}$, the effect of the SOI constant $\alpha_{SO}$ is not strong; with increasing $\alpha_{SO}$, $R_{++}$ slightly increases as it is expected from the semi-classical theory. Numerically, the quasi-classical regime of reflection can be obtained for strong SOIs and smooth barriers. In fig. 6, we show the functions $R_{\alpha,\alpha'}(a)$ for the case of $\alpha_{SO} = 10^{-5} meV\, cm$, which can exist in other narrow-band semiconductors. It is seen that the off-diagonal reflection coefficients strongly decrease with increasing the width $a$, whereas the diagonal coefficients approach unity. For $a = 10000\,\overset{\circ}{A}$, the parameter $\eta \approx 0.36 < 1$ and the spin-orbit motion of electron becomes adiabatic.

**Discussion**

An observation of the predicted multi-beam reflection depends on two factors. On one hand, one needs a sufficiently strong SOI, which would result in strong spatial separation of beams with different spin polarizations (figs. 1 and 2c). On the other hand, typical lateral dimensions of etched meso-scopic structures are in the sub-μm range and, if the SOI is very strong, the spin-conversion efficiencies ($R_{+\to-}$ and $R_{-\to+}$) can become small. Here we found that the InSb quantum wells would be a suitable system to observe this effect. InSb quantum wells have the moderate SOI, which results in the spin-dependent angular deviations, $|\theta_{+\to-} - \theta|$ and $|\theta_{-\to+} - \theta|$, of order of $10°$ [3]. At the same time, the off-diagonal reflection coefficients for relatively smooth barriers remain large. Another suitable system to observe spin-dependent reflection can be a 2D hole gas in GaAs quantum wells [5,12] where the SOI is quite strong due to the mixing between heavy and light holes.

To conclude, we have studied the physical properties of reflection of electrons from a lateral barrier in a narrow-gap semiconductor. Using the typical parameters of InSb quantum wells, we show that the effect of multi-beam reflection remains strong for realistic lithographical barriers. The spin-dependent reflection described in this paper can be used for spin filtering devices based on ballistic nano-structures.

**Acknowledgments.** A.O.G would like to thank  I. V. Zozoulenko, J. Heremans, and M. Santos for helpful discussions. This work was supported by Ohio University, and by Volkswagen and A.v.H. Foundations.

**Figure captions**

Fig.1. Two geometries utilizing ballistic electron beams and multi-beam reflection from a barrier. Electron trajectories are calculated from the quasi-classical equations in the presence of a weak normal magnetic field. The angles of reflected waves are determined by the kinematics equations. The 2D density is $3 \cdot 10^{11} \, cm^{-2}$ and $E_F \approx 51 \, meV$; the SOI parameters correspond to the InSb quantum wells.

Fig. 2. a) Two Fermi circles in a 2D system with the SOI. b) Geometry of multi-beam scattering. c) Scattering angles as a function of the incident angle; $\alpha_{SO} = 10^{-6} \, meV \, cm$. The 2D density is $5 \cdot 10^{11} \, cm^{-2}$ and $E_F \approx 85 \, meV$. Inserts: geometries of scattering for differing incoming waves.

Fig. 3. Different types of electron trajectories in the presence of a smooth potential barrier (a,b,c). d) Sketch of the barrier.

Fig. 4. Diagonal and off-diagonal reflection coefficients as a function of the incident angle. The barrier heights are 150 *meV* (a) and 70 *meV* (b). The 2D density is $2.1 \cdot 10^{11} \, cm^{-2}$ and $E_F = 35 \, meV$.

Fig. 5. Reflection coefficients as a function of the incident angle for different SOI constants. The barrier height is $150\, meV$, $N_{2D} = 2.1 \cdot 10^{11}\, cm^{-2}$, and $E_F = 35\, meV$.

Fig. 6. Reflection coefficients as a function of the barrier width for the stronger SOI, $\alpha_{SO} = 10^{-5}\, meV\, cm$. The 2D density is $2.1 \cdot 10^{11}\, cm^{-2}$ and $E_F = 35\, meV$.

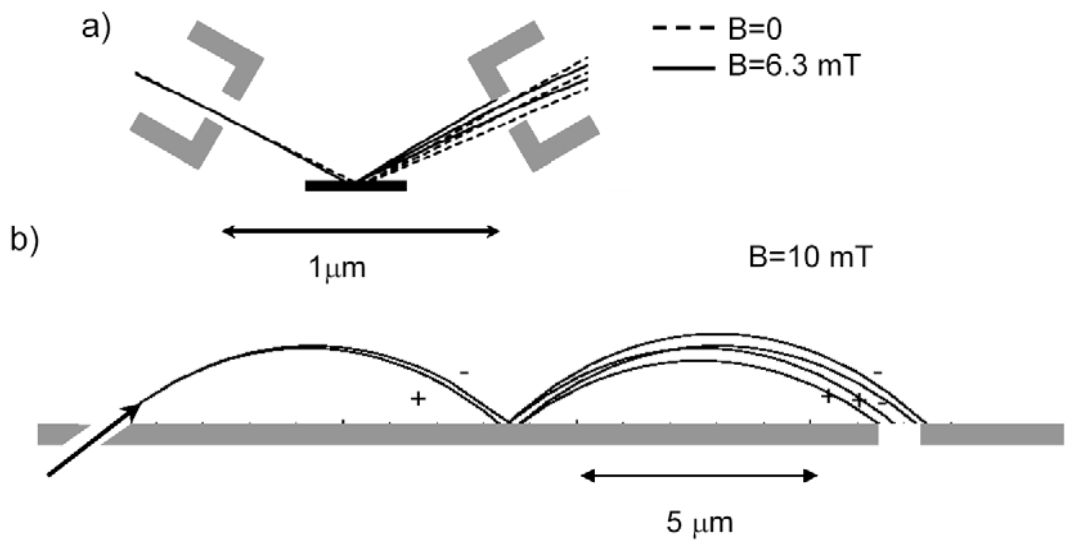

Figure 1, Govorov et al.

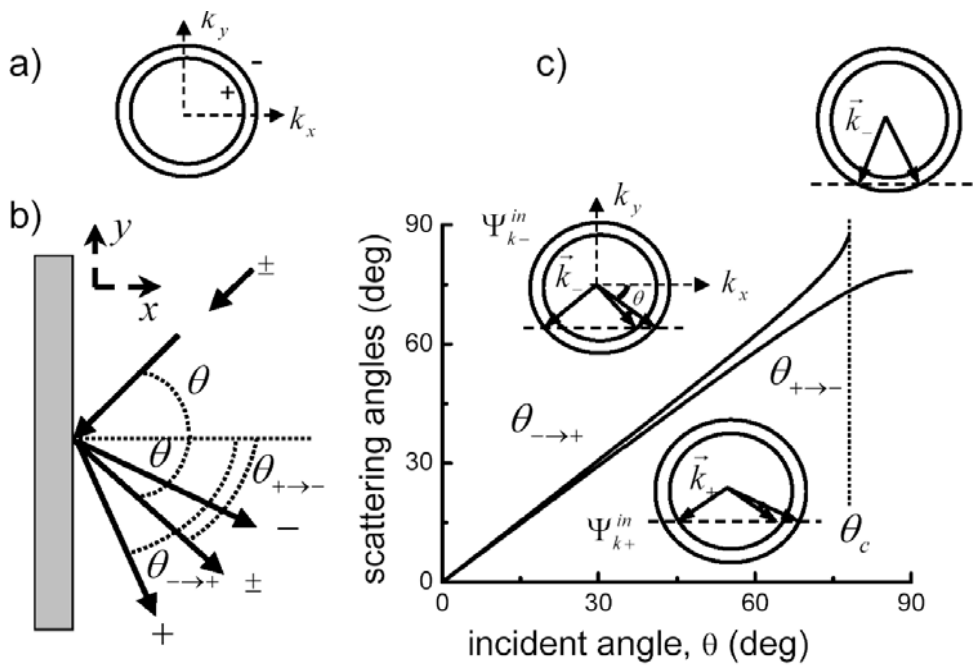

Figure 2, Govorov et al.

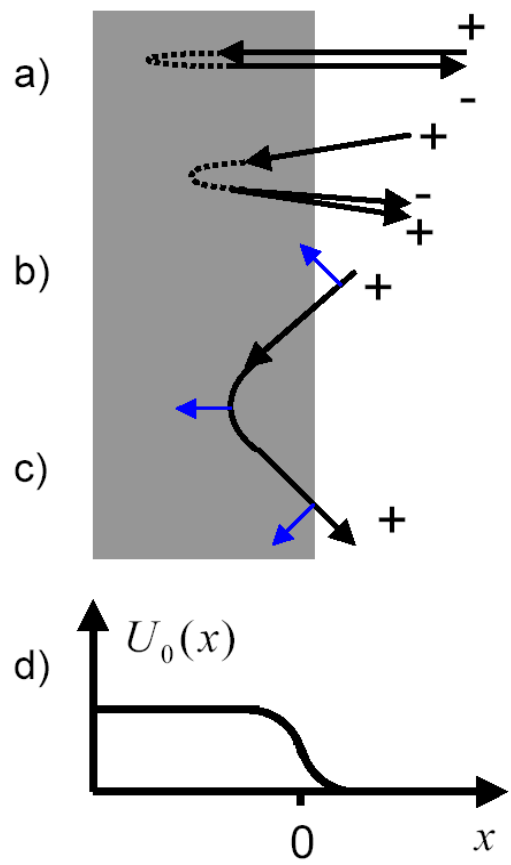

Figure 3, Govorov et al.

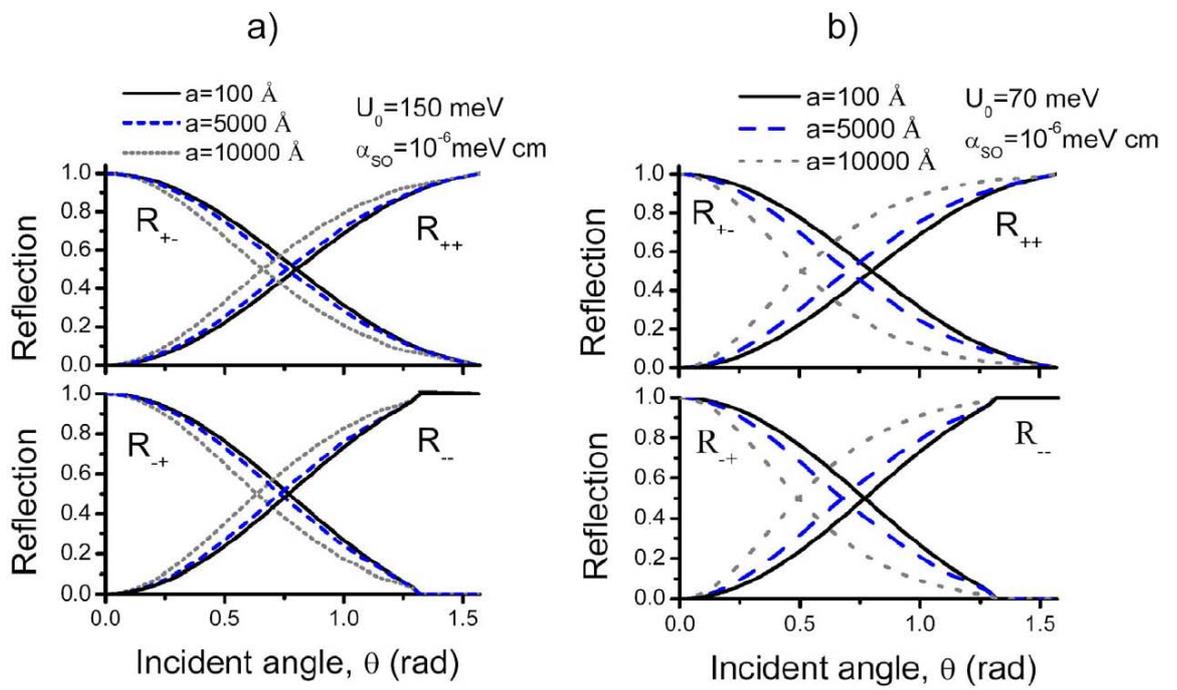

Figure 4, Govorov et al.

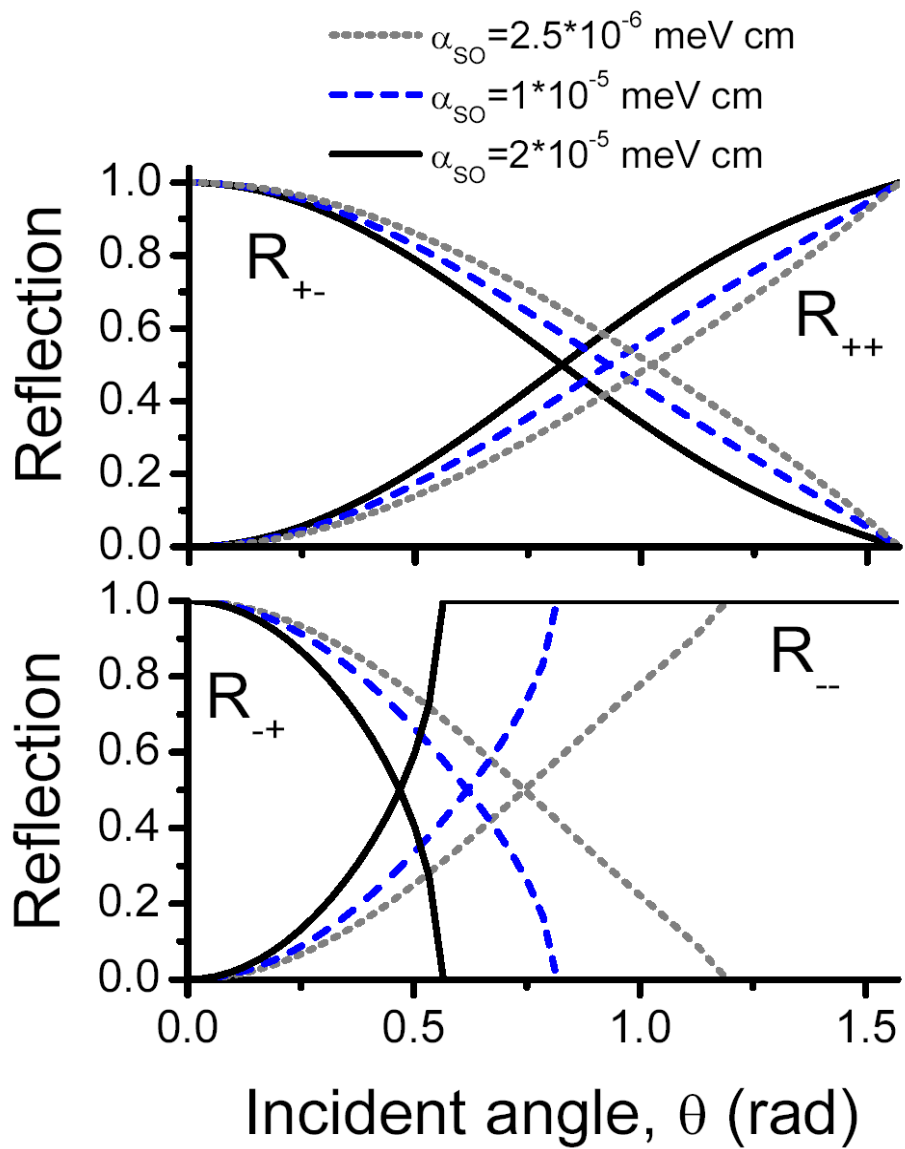

Figure 5, Govorov et al.

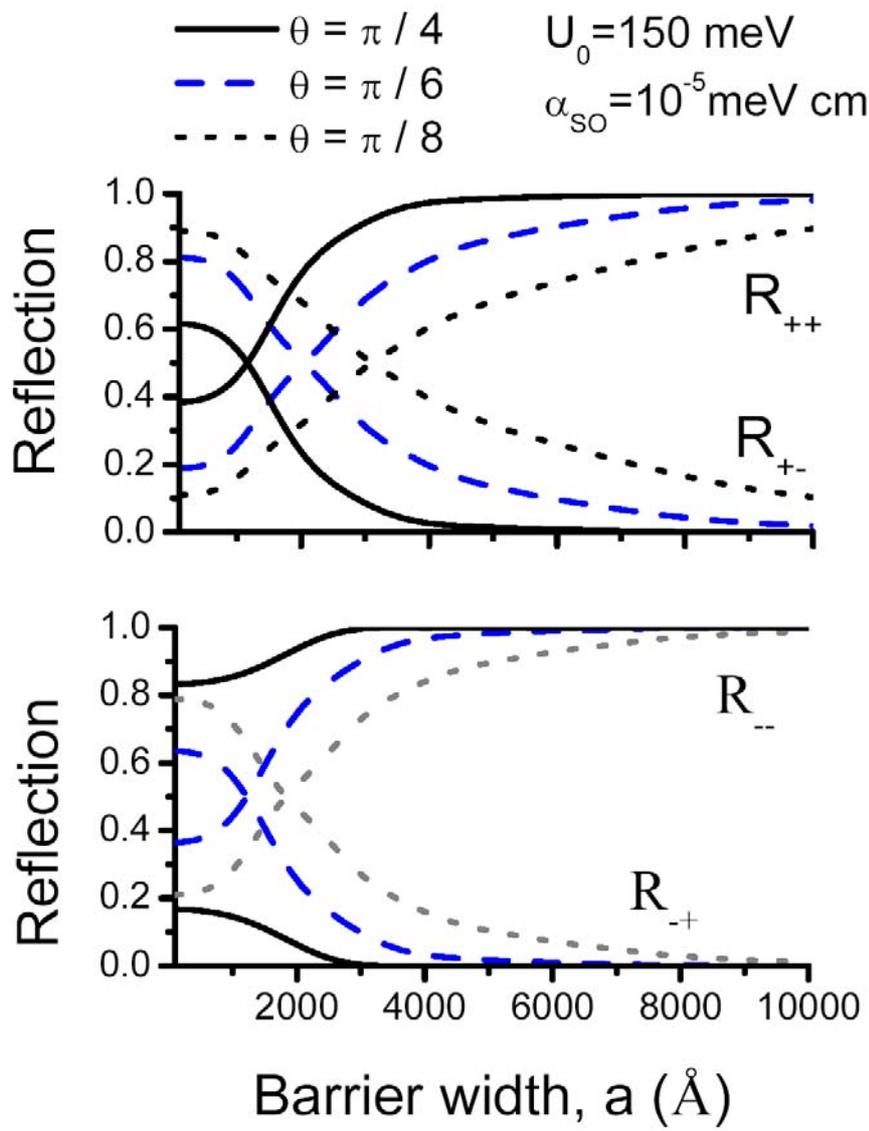

Figure 6, Govorov et al.